# *A mid-infrared dual-comb spectrometer in step-sweep mode for high-resolution molecular spectroscopy*

Muriel Lepère*[a], Olivier Browet[a], Jean Clément[a], Bastien Vispoel[a], Pitt Allmendinger[b], Jakob Hayden[b], Florian Eigenmann[b], Andreas Hugi[b], and Markus Mangold[b],

[a]Research unit Lasers and Spectroscopies (LLS), Institute of Life, Earth and Environment (ILEE), University of Namur, 61, Rue de Bruxelles, Namur, Belgium;
[b]IRsweep AG, Laubisrütistrasse 44, 8712 Stäfa, Switzerland

*muriel.lepere@unamur.be; phone +3281724496; https://www.unamur.be/en/sci/physics/ur-en/lls/

ABSTRACT

To meet the challenges of high-resolution molecular spectroscopy, increasingly sophisticated spectroscopic techniques were developed. For a long time FTIR and laser-based spectroscopies were used for these studies. The recent development of dual-comb spectroscopy at high-resolution makes this technique a powerful tool for gas phase studies. We report on the use and characterization of the IRis-F1, a tabletop mid-infrared dual-comb spectrometer, in the newly developed *step-sweep* mode. The resolution of the wavenumber axis is increased by step-wise tuning (interleaving) and accurate measurement of the laser center wavelength and repetition frequency. Doppler limited measurements of $N_2O$ and $CH_4$ reveal a wavenumber accuracy of $10^{-4}$ cm$^{-1}$ on the covered range of > 50 cm$^{-1}$. Measured half-widths of absorption lines show no systematic broadening, indicating a negligible instrument response function. Finally, measurements of nitrogen pressure broadening coefficients in the $\nu_4$ band of methane show that quantum cascade laser dual-comb spectroscopy in *step-sweep* mode is well adapted for measurements of precision spectroscopic data, in particular line shape parameters.

## 1. INTRODUCTION

High-resolution molecular spectroscopy has both fundamental and applied interests. It allows understanding intra- and inter-molecular interactions in gas phase, and is a powerful tool to study planetary atmospheres. On Earth, atmospheric pollution and global warming are challenges for all societies around the world [1, 2]. Different instruments installed on various platforms (space satellites, stratospheric balloons, ground-based stations) are continuously monitoring the atmosphere to characterize it and unravel the mechanisms occurring. The retrieval of atmospheric spectra relies on the computation of the radiative transfer that in turn requires many chemical and physical parameters, among them the spectroscopic parameters and their temperature evolution. The recent improvements of remote sensing mission instruments have pushed the need for high accuracy spectroscopic data [3, 4], as the quality of the retrieved information depends on the precision of the spectroscopic parameters [5] for which the line-shape parameters are the largest sources of uncertainties [6-8].

High-resolution infrared spectrometers, such as high-resolution dual-comb spectrometers, are able to provide precise spectroscopic parameters. This data is also very important for the development and improvement of theoretical models describing molecular interactions. For decades, theoretical models have been developed and used [9] to describe the molecular interactions and to compute line-shape parameters. The models rely on intermolecular potentials that must be validated by accurate measurements. Theoretical models and laboratory measurements complement, challenge and boost each other in this quest to understand chemical and physical phenomena that occur in the gas phase. In particular, the mid-infrared spectral region is very interesting since the fundamental vibrational bands of many molecules are located in this spectral domain. At room temperature, these bands are the most intense; this allows a great precision in the measurements of spectroscopic data.

To meet the challenges of molecular spectroscopy, increasingly sophisticated spectroscopic techniques have been developed. They compete in a variety of often conflicting parameters such as good spectral resolution and large coverage, accurate absolute frequency calibration, and high signal-to-noise ratio at short measurement times. For a long time, Fourier transform spectroscopy (Ref. [10] and therein) was the workhorse for mid-infrared studies providing a lot of accurate line parameters. In addition to a large number of commercial FTIR spectrometers, the spectral resolution has been pushed with specialized instruments with an optical path difference of up to 22 m [11] (see Ref. [12] and therein). More recently, laser-based spectroscopies have made a big impact on high-resolution spectroscopy (Ref. [12] and references in it). While sources such as diode lasers or single mode quantum cascade lasers (QCL) limit the spectral range to a few wavenumbers at most, they provide great spectral resolution, high signal-to-noise ratio, and extremely short measurement times [13, 14]. These techniques have further been coupled with technologies such as cavity ring down [15, 16], allowing the measurement of gases in extremely low concentration, cold gases in molecular jets [17, 18], or specialized cells [19-23] that enable the study of line parameters over a large temperature range.

The advent of frequency combs has revolutionized the field of high-resolution molecular spectroscopy [24]. In the near-infrared spectral range, many different spectroscopic techniques have been developed which can roughly be divided in comb-assisted spectroscopy [25] and direct frequency comb spectroscopy [26-31]. Dual-frequency comb spectroscopy is a type of the latter, which got much attention because of its mechanical simplicity (no moving elements) and the resulting potential for fast measurements of full spectra [32-36]. The mid-infrared spectral range – of interest because it hosts the fundamental ro-vibrational transitions of many molecules – has proven technologically more challenging because direct frequency comb sources were not available for a long time. Comb-assisted techniques have pushed the spectral resolution of single-mode QCLs to their limit [37, 38], while direct frequency comb spectroscopy relied on non-linear conversion of near-infrared combs in optical parametric oscillators (OPOs) or by difference frequency generation (DFG) [39, 40].

Only more recently, chip-scale sources of mid-infrared frequency combs were demonstrated in micro-resonators [41], inter band cascade lasers (ICL) [42, 43] and quantum cascade lasers [44]. Chip-scale frequency comb sources bear some inherent advantages for molecular spectroscopy. Their small size in combination with their relatively low power consumption makes them particularly attractive for field-deployed sensors or mobile applications [43, 45, 46]. The large optical output power distributed on a rather low number of optical modes bears the potential for non-linear or saturated spectroscopy [47]. In dual-comb implementations, the large repetition frequency inherent to small-scale sources allows for rapid acquisition of the full spectrum covered by the combs [48-52]. Conversely, the large repetition frequency leads to a sparse sampling of the spectrum impeding the use for high-resolution spectroscopy. Interleaving – also known from other comb sources [53-55] – has been demonstrated in micro-resonators [56-58] and ICLs [42]. MHz-level resolution and frequency accuracy in the mid-infrared has been demonstrated with QCL frequency combs. Gianella and co-workers demonstrated so called *rapid-sweep* interleaving by applying synchronized current ramps to tune both combs simultaneously [59].

In a joint effort between IRsweep and the University of Namur, we have developed for the first time what we call a *step-sweep* approach to mid-infrared high-resolution dual-comb spectroscopy with QCLs, building on a method for spectral interleaving described by Villares, Hugi, Blaser and Faist [60]. The presented technique relies on step-by-step tuning of the lasers. In contrast to simultaneous tuning of the lasers, the demonstrated technique allows for direct measurement of the relative frequency axis from the heterodyne beat signal without relying on either high bandwidth locking loops or external frequency rulers, except for the clock of the acquisition card. Furthermore, *step-sweep* retains the microsecond time resolution of the dual-comb measurement making it applicable to studies of reacting gases or transient measurement of pulsed molecular beams [61]. For an absolute frequency calibration, two parameters (an overall offset and line spacing) need to be retrieved from the measurement of a known calibration substance. Despite the optical and electronic simplicity of the presented technique, we demonstrate, over the complete coverage of > 50 cm$^{-1}$, a frequency accuracy of below 12 MHz (0.0004 cm$^{-1}$) in separately calibrated measurements and below 4 MHz (0.00012 cm$^{-1}$) when the calibration is simultaneous with the sample measurement. Furthermore, the line broadening due to the instrument response is found to be negligible for Doppler broadened absorption lines and the line width is determined with an inaccuracy of < 1.5 MHz (0.00005 cm$^{-1}$). With these characteristics, the here demonstrated *step-sweep* technique can be employed for challenging tasks in molecular spectroscopy such as measurement of line shape parameters, which is demonstrated by the study of the $N_2$-collisional broadening coefficient of methane lines in the $\nu_4$ fundamental band. We show that the agreement with literature is very good.

## 2. IRIS-F1, A MID-INFRARED DUAL-COMB SPECTROMETER

### 2.1 Principle

High-resolution spectra were recorded using a commercial quantum cascade laser dual-comb spectrometer (IRis-F1, IRsweep AG) that was customized with the *step-sweep* technology for wavelength tuning (compare section 2.2). Dual-comb spectroscopy is a type of Fourier transform spectroscopy, where the mechanical interferometer is replaced by the beating of two multi-mode coherent light sources [26, 40], which we call the local oscillator and the sample comb. In brief: each frequency comb light source emits a spectrum consisting of multiple, spectrally narrow emission frequencies. In a given comb, all the emitted frequencies have a constant spacing. Each emission frequency of the local oscillator comb (green lines in top panel of figure 1) has a close-by emission frequency of the sample comb (blue lines in figure 1). Upon mixing on a detector, a beat signal at the difference frequency of the two lines is generated. Because of a slightly different spacing between the lines of the local oscillator and the sample comb, each pair of lines generates a beat signal at a distinct frequency. Note that, since only the sample comb passes through the sample, the attenuation observed on the radio frequency signal corresponds to half of the absorbance of the sample gas experienced by the sample comb (compare figure 1b).

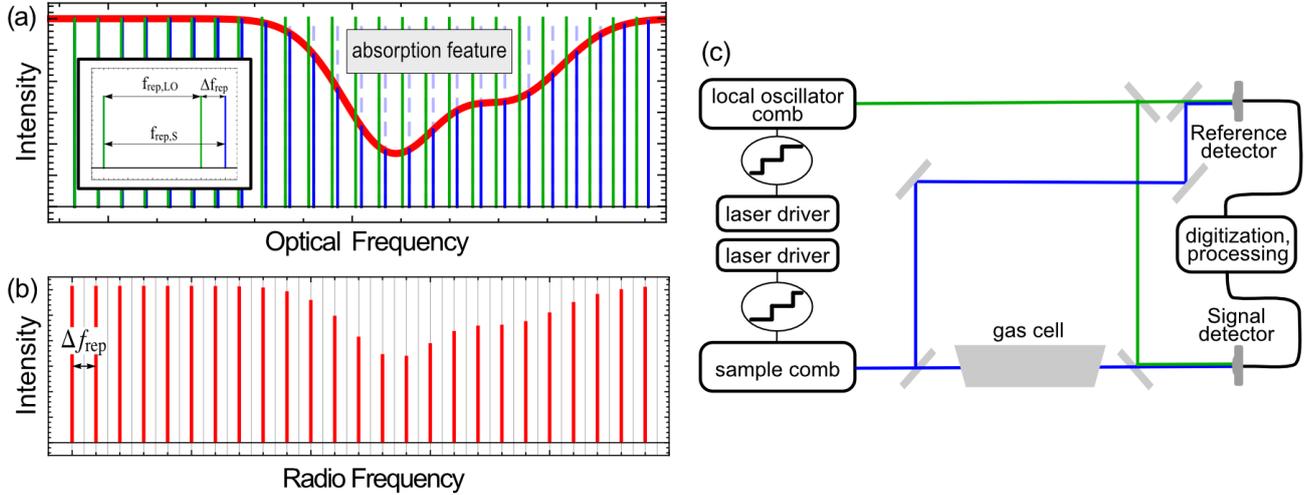

Figure 1. Left: Illustration of the optical spectrum of the sample (blue) and reference comb (green) (top panel) and the multiheterodyne spectrum measured on a detector in DCS (bottom panel). Right: Functional schematic of the IRis-F1 dual-comb spectrometer in phase sensitive configuration

A schematic of the spectrometer is shown in Figure 1. The spectrometer uses two closely matched QCL frequency comb sources emitting at a center frequency of 1308 cm$^{-1}$ with a span of 50 cm$^{-1}$ and a repetition rate of 9.891 GHz (0.3299 cm$^{-1}$) and 9.894 GHz, respectively. The spectrometer is operated in a phase sensitive configuration [59], *i.e.* only one of the beams (sample comb) passes through the sample and is overlapped with the beam from the second QCL frequency comb (local oscillator) before being detected on a thermoelectrically cooled HgCdTe detector of 1 GHz bandwidth. The detector signal displaying a multi-heterodyne beat note between the two frequency combs is digitized using a high speed digitizer (2 GS, 14 bit). The raw digitized signal is processed on dedicated hardware as described by Klocke *et al.* [48] to obtain the amplitude and phase of each line in the heterodyne beat note, corresponding to the intensity and phase of each line of the sample QCL frequency comb. Since the frequency combs are free running, i.e. they are not locked to one another, small frequency fluctuations result in amplitude and frequency noise on the recorded multiheterodyne beat note. This noise is strongly reduced using a reference beam-path and detector (compare Figure 1) and calculating the ratios of the complex sample and reference beat note amplitudes [48].

### 2.2 Step sweeping for high resolution

Although the narrow line width of the individual lines of the QCL frequency comb is well suited for high-resolution spectroscopy, the point spacing between adjacent lines is generally too large. While the high repetition rate $f_{rep}$ of QCL frequency combs (0.3299 cm$^{-1}$ in this work) and large $\Delta f_{rep}$ of the IRis-F1 dual-comb spectrometer render it ideally suited for µs-time-resolved spectroscopy in condensed phase [48], narrow features of molecular gases are sampled insufficiently by the lines of the comb. To close the gaps between adjacent comb lines, the QCL frequency combs' emission frequencies can be tuned across a gap via temperature or current. Gianella *et al.* used a synchronized ramp of the sample and local oscillator comb's wavelengths to keep the difference frequencies between them, and hence the multiheterodyne beat note, within the detection bandwidth of the photo detectors [59]. Since the wavelength of both lasers were changed simultaneously, the wavelength and repetition rate tuning of the frequency combs cannot be retrieved from the multiheterodyne beat note. Hence, the wavenumber axis must be retrieved from reference data, e.g. the interference fringes of an etalon, or by including frequency references, such as line-locked single-mode QCLs, in the experiment [62].

Here, we overcome this limitation by tuning only one QCL frequency comb at a time, using the second frequency comb as a fixed reference. The procedure is illustrated in figure 2. Initially, the optical spectrum of the local oscillator (LO) and signal (S) frequency comb are described by the amplitudes and phases of the comb lines of index *j* centered at frequencies

$$f_{j,LO} = f_{CEO,LO} + j \cdot f_{rep,LO}$$

$$f_{j,S} = f_{CEO,S} + j \cdot f_{rep,S}$$

Herein, $f_{CEO}$ is the center frequency of the comb and $f_{rep}$ the line spacing (repetition frequency) of the comb. For the combs used here, the index j covers roughly the range $-90 < j < 90$. The frequencies $f_{j,het}$ of the lines of the multiheterodyne spectrum measured on the detectors are described in the same way by a center frequency $\Delta f_{CEO}$ and repetition frequency $\Delta f_{rep}$ given by

$$f_{j,het} = \Delta f_{CEO} + j \cdot \Delta f_{rep}$$

$$\Delta f_{CEO} = f_{CEO,S} - f_{CEO,LO}$$

$$\Delta f_{rep} = f_{rep,S} - f_{rep,LO}$$

The amplitude R and phase φ of the lines in the multiheterodyne spectrum (used for calculating transmission and dispersion spectra) are measured relative to the multiheterodyne spectrum from a reference detector [48].

A step along the frequency axis is initiated by increasing the current of the signal comb by an increment, yielding a shift of $f_{CEO,S}$ and $f_{rep,S}$ and a corresponding change of $\Delta f_{CEO}$ and $\Delta f_{rep}$ (not shown in Figure 2) of the multiheterodyne spectrum. Since the local oscillator FC is unchanged, the shift of each line in the multiheterodyne spectrum corresponds directly to the shift of the optical frequency of the signal FC.

$$f_{CEO,S,k}(k = 1, 3, \dots) = f_{CEO,LO,k} + \Delta f_{CEO,k}$$

$$f_{rep,S,k}(k = 1, 3, \dots) = f_{rep,LO,k} + \Delta f_{rep,k}$$

$$f_{CEO,LO,k}(k = 1, 3, \dots) = f_{CEO,LO,k-1}$$

$$f_{rep,LO,k}(k = 1, 3, \dots) = f_{rep,LO,k-1}$$

where, *k* is the index of the current step. Odd and even values of *k* correspond to steps of the signal frequency comb and local oscillator frequency comb, respectively. After stabilization of the signal frequency comb and measurement of $\Delta f_{CEO,k}$ and $\Delta f_{rep,k}$, the local oscillator frequency comb is tuned in the same way. Thereby, $\Delta f_{CEO,k}$ and $\Delta f_{rep,k}$ of the multiheterodyne spectrum are shifted back close to their initial values. Again, since only the local oscillator frequency comb changes, the measured values $\Delta f_{CEO,k}$ and $\Delta f_{rep,k}$ in the beat note directly reflect the changes of the optical frequencies of local oscillator comb and the new $f_{CEO,LO,k}$ and $f_{rep,LO,k}$ are given by

$$f_{CEO,LO,k}(k = 2, 4, \dots) = f_{CEO,S,k} - \Delta f_{CEO,k}$$

$$f_{rep,LO,k}(k = 2, 4, \dots) = f_{rep,LO,k} - \Delta f_{rep,k}$$

$$f_{CEO,S,k}(k = 2, 4, \dots) = f_{CEO,S,k-1}$$

$$f_{rep,S,k}(k = 2, 4, \dots) = f_{rep,S,k-1}$$

By repeating the current steps on the signal and local oscillator frequency comb, the gap between adjacent lines can be closed without moving the multiheterodyne beat note to frequencies outside the detection bandwidth. Since $\Delta f_{CEO}$ and $\Delta f_{rep}$ are measured in every step, only the initial values of $f_{CEO,S,k=0}$ and $f_{rep,S,k=0}$ are necessary to precisely locate each measurement on the frequency axis. Here, $f_{CEO,S,k=0}$ and $f_{rep,S,k=0}$ are determined in post-processing by fitting the recorded absorption line positions to reference data from published line lists.

In conclusion, the relative wavenumber scale can be obtained directly from each measurement via the heterodyne beat frequencies. However, we found that the repeatability of the wavenumber axis between scans was better than its measurement due to the accumulation of measurement errors over many steps. Therefore, the relative wavenumber scale is obtained from a series of sweeps by co-averaging the step sizes and is then used for all subsequent measurements taken with the same laser conditions. After that, the absolute frequencies of one step of the sweep – given by the two parameters $f_{CEO,S,k=0}$ and $f_{rep,S,k=0}$ – have to be determined from at least two known absorption features of a sample. While best accuracy is achieved by measuring the calibration substance simultaneously with the sample of interest, we will show in section 3.2, that the long term stability of the frequency combs also allows determining the initial offset and repetition frequency in a separate measurement before or after the actual sample measurement.

To obtain transmission and dispersion spectra, a background spectrum is recorded by repeating the above procedure while the sample gas cell is evacuated (background spectrum) and filled with the sample (sample spectrum).

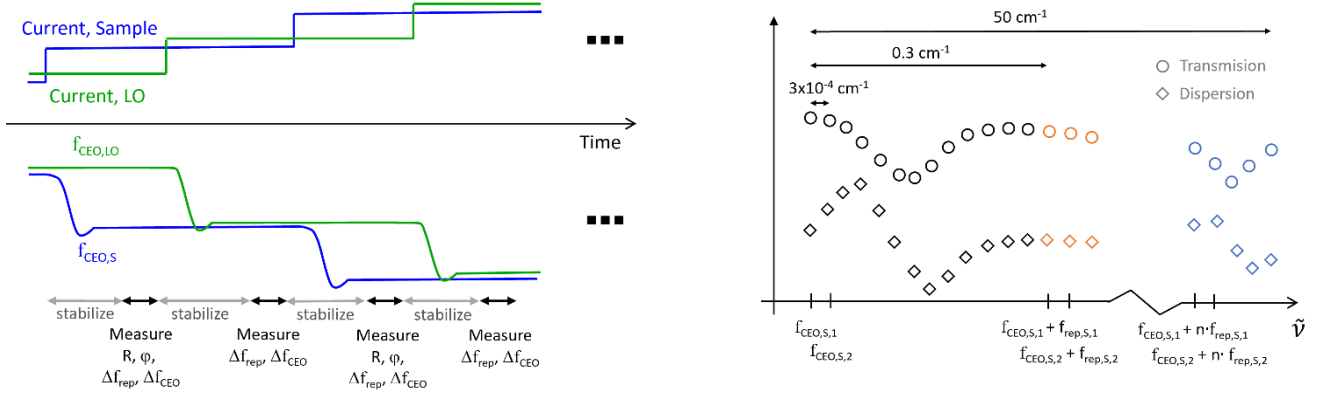

*Figure 2. Illustration of the step sweep technique (see text). Left: Laser current and comb center frequency of the signal (S) and local oscillator (LO) comb during step sweeping. Right: Schematic transmission and dispersion spectra. Different colors indicate different comb lines that are tuned in a step-wise fashion to bridge the gap to the next comb line.*

### 2.3 Experimental details

The step width can be varied between 6 MHz (0.0002 cm$^{-1}$) and approximately 200 MHz (0.0067 cm$^{-1}$), corresponding to 1700 to 50 steps to fill the gap between adjacent comb lines. The lower step width limit is given by the resolution of the digital to analog converter that is used to adjust the current setpoint on the laser driver. The upper limit is given by the bandwidth of the data acquisition card of 1 GHz, of which ~ 500 MHz is occupied by the width of the multiheterodyne beat-signal (~ 180 lines, $\Delta f_{rep}$ = 3.7 MHz) and 0-50 MHz and 950 MHz – 1 GHz are typically avoided due to low frequency noise and the onset of the high-frequency roll-off, respectively.

The duration of a wavelength sweep depends on the number of steps (~1.1 · $f_{rep}$ / step-width) and the target signal to noise ratio (SNR) on the vertical axis (transmission and dispersion). The SNR increases with the square root of the acquisition length in the range between 4 µs and ~ 125 ms. In this work, 4.2 ms were used (compare section 3.1). The stabilization time of 1 s after each current step is larger than the data processing time for the acquisition. The overall measurement time of 1200 s per step sweep measurement of 2x600 step is hence limited by the laser stabilization time. For acquisition lengths exceeding 4.2 ms, the data processing time becomes limiting.

## 3. CHARACTERISATION OF IRIS-F1 AT HIGH RESOLUTION

### 3.1 Spectral range, absorbance noise, and linear range

The spectral coverage in a dual-comb experiment is given by the overlap region of the two used frequency combs. The lasers used in this study overlap in a spectral range from 1283 to 1333 cm$^{-1}$. This range is well suited to study the Q-branch as well as the low rotational levels of the P- and R-branches of the $v_4$ fundamental band of methane. In Figure 3 a), a measured transmission spectrum of pure methane at 0.185 mbar pressure is presented together with a calculated transmission spectrum based on line parameters from the HITRAN database [63]. Figure 3 b) shows the dispersion spectrum which is measured simultaneously in the dual-comb experiment.

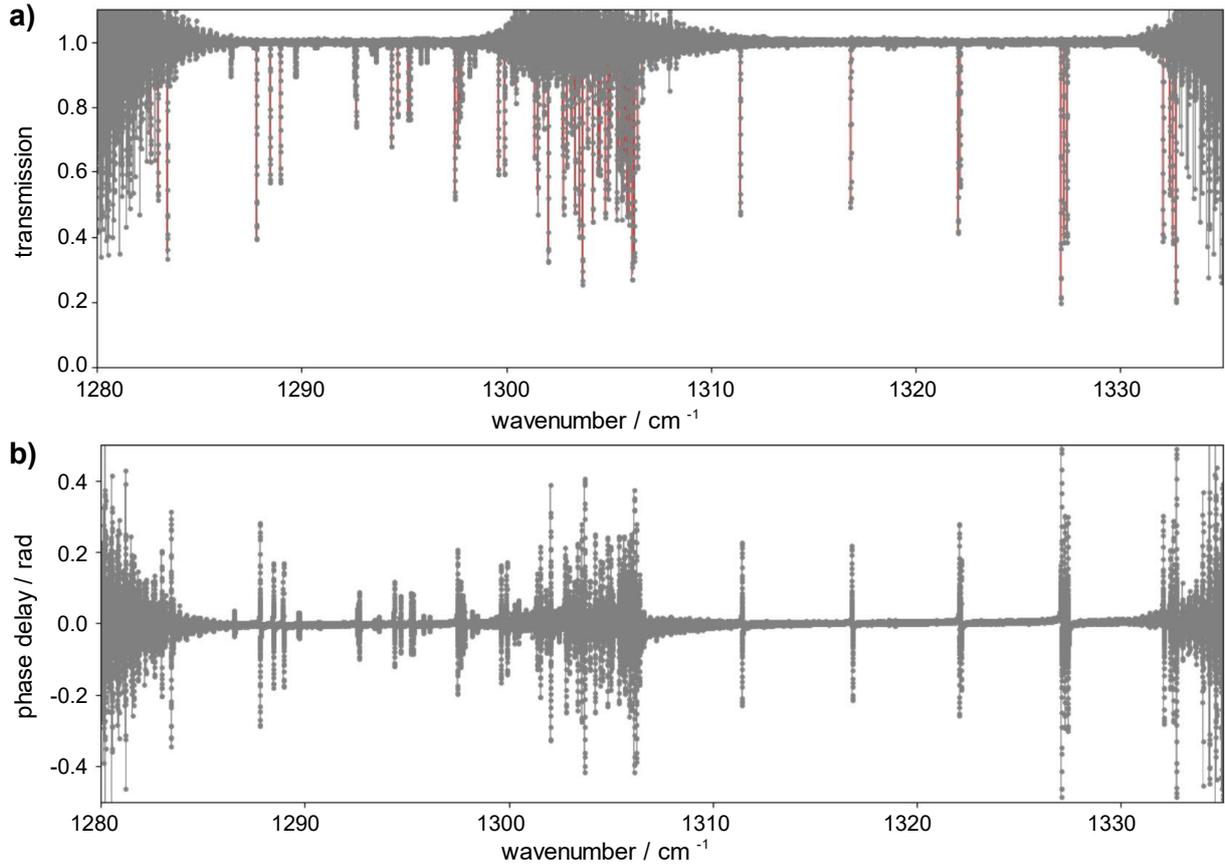

*Figure 3: a) Transmission and b) corresponding dispersion spectrum of room temperature, Doppler broadened methane at 0.185 mbar pressure. The red line represents a calculated transmission spectrum based on Hitran line parameters.*

Since the emission power varies from comb mode to comb mode (compare inset in Figure 4 b), the signal quality varies considerably within the covered spectral range. In Figure 4 a), we plot the absorbance $A = -\ln(T)$ of a transmission $T$ measurement of an empty gas cell (grey dots). The red dots show a co-average of five subsequent measurements. We observe two low noise regions interrupted by a higher noise region covering roughly 1300-1308 cm$^{-1}$. The inset shows a zoom on the absorbance signal around 1292 cm$^{-1}$. A more comprehensive view of the same data is given in Figure 4 b), where for each comb mode the standard deviation of the 600 measurement steps is plotted vs. their center wavenumber. Again, the grey solid dots are values obtained with a single scan, while the red ones are computed after co-averaging five measurements. For a single scan, the standard deviation lies below 0.01 absorbance units on a range of 36 cm$^{-1}$, while in the complete range used for analysis in this paper from 1283 to 1333 cm$^{-1}$ the standard deviation lies below 0.05 absorbance units. The signal quality can be improved by co-averaging approximately by the expected factor of $\sqrt{5}$ as shown by the red dots. Also, the high noise region in the center could be further suppressed by using spectral filters to shape the laser emission spectrum [64].

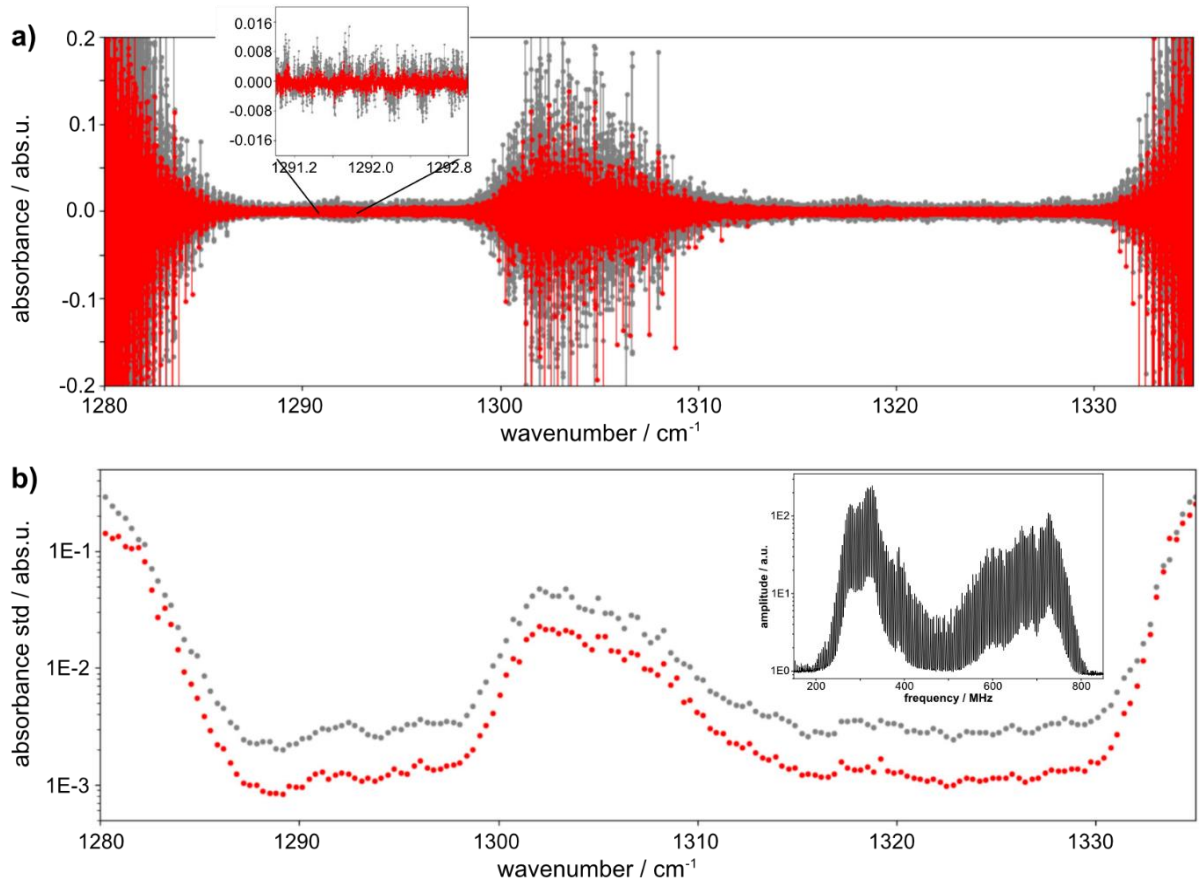

*Figure 4: a) Absorbance measurement of an empty cell. Inset: zoom on the range 1291-1293 cm-1. Grey dots result from a single scan, red dots are 5 scans co-averaged. b) Each grey (red) data point represents the standard deviation of the absorbance signal of a single comb mode calculated over all 600 steps for a single scan (5 scans co-averaged). Inset: Heterodyne emission spectrum of the dual-comb laser module.*

To investigate the linear range of the dual-comb measurement and the performance at very high optical density (strong absorbance), the spectrum of $CH_4$ at a pressure of 5.06 mbar and at room temperature was recorded. In figure 5a), a section of the obtained absorption spectrum is plotted together with a reference spectrum calculated using line parameters from the HITRAN 2020 database [63].

The peak absorbance of lines from 1283 to 1333 cm$^{-1}$ were extracted by means of peak finding without fitting of the absorption band. The found values were compared against the peak absorbance predicted by the HITRAN 2020 simulation. The result is shown in figure 5b) for absorption lines exceeding a peak absorbance of 0.006. Agreement with the HITRAN reference spectrum is observed for absorption lines of peak absorbance in the entire range from the measurement noise floor (compare figure 4 b) to 10 abs.u., limited by dark noise of the detection system. The ability to accurately measure very strong absorbance can be attributed to the phase coherent dual-comb measurement, which reduces the effect of incoherent dark noise of the detection system.

The linearity and large dynamic range of the transmission axis indicate that the spectrometer is well suited for measurements of line intensities. However, such measurements are beyond the scope of this article since a shorter gas cell of precisely defined length as well as a higher sample pressure would be needed.

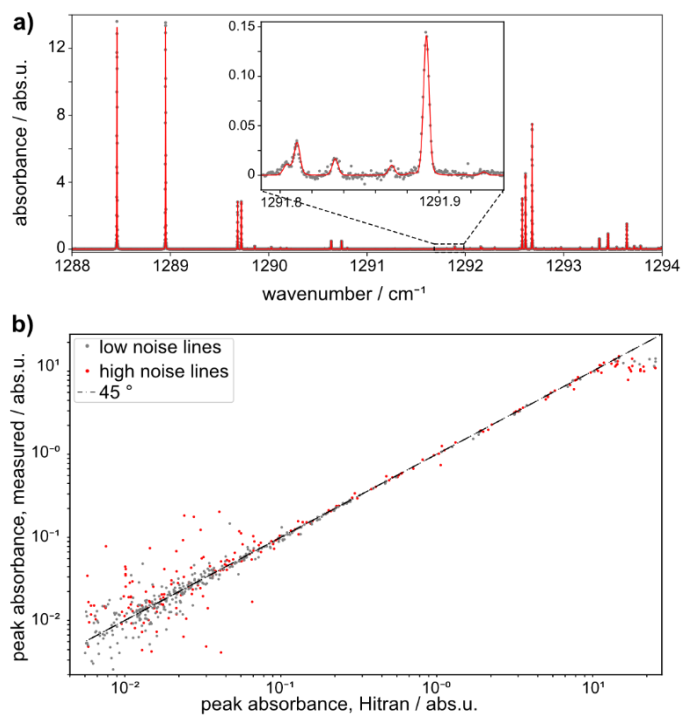

*Figure 5. Linear range of absorption measurements. a) Measured (grey dots) and Hitran simulated spectrum (red line) of CH4 at 5.06 mbar, featuring strong absorption lines. b) From the same measurement, measured peak absorbance of lines from 1283 to 1333 cm-1 plotted against their predicted value. Data points are categorized into spectral regions with high and low measurement noise.*

### 3.2 Wavenumber calibration of spectra

As described in section 2.2, the wavenumber axis calibration of the spectrometer consists of two steps: (i) creating a relative wavenumber axis from the measurement of the heterodyne beat frequencies of "empty" sweeps and (ii) determining the starting frequencies $f_{CEO,S,k=0}$ and $f_{rep,S,k=0}$ from the measurement of a well-known calibration spectrum. To assess the accuracy and repeatability of the relative wavenumber axis and the starting frequencies, we analyze measurements of the $v_1$ band of N$_2$O and the $v_4$ band of methane and compare them to tabulated values.

For the two considered molecules, the absorbance of each line in the spectrum has been fitted using a Gaussian profile (the partial pressure of active gas was respectively 0.045 mbar and 0.182 mbar for N$_2$O and CH$_4$, allowing us to neglect the contribution of collisions). In figure 6, we show a fit of a Gaussian profile to an absorption line of methane at 1288.457 cm$^{-1}$ in red. The standard deviation of the fit residual of 0.0019 absorbance units corresponds well to the standard deviation determined in the empty cell measurements shown in figure 4 b). We generally find that the residuals of Gaussian fits to low pressure methane transitions correspond to values expected from the measurement noise.

A fit of the corresponding dispersion signal with a Dawson function is shown in black in figure 6. Both, dispersion and absorbance signal, can well be used to determine the line position. We generally observe lower fit uncertainties in the dispersion signal in the high noise part of the spectrum. At the same time, the dispersion signal is more strongly influenced by other transitions close to the fitted transition. Therefore, it should only be used on well isolated transitions or multiple transitions have to be fitted simultaneously. To avoid these complications, for the rest of this study we only analyze results obtained by fitting the absorption signal.

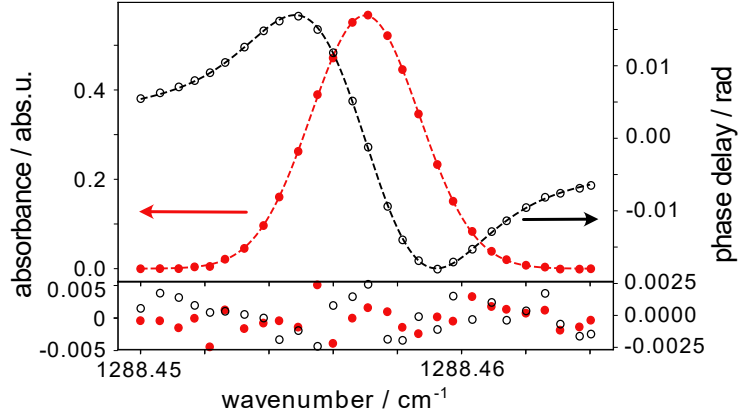

*Figure 6. Top panel: absorbance (red dots) and dispersion (black circles) corresponding to the transmission and dispersion data shown in Figure 3, zoomed in around 1288.457 cm-1. The red and black dashed lines indicate a fitted Gaussian and Dawson profile, respectively. Bottom panel: residuals of the fitted line shapes.*

The quality of the relative frequency axis is assessed by comparing the observed line frequencies of the $N_2O$ transitions to the HITRAN line positions [63]. Two recent experiments have confirmed the line positions given in Hitran to within < 2 MHz deviation [65, 66]. Comparing our data to these line lists leads to the same conclusion, therefore we restrict our comparison to the values given in Hitran. For this comparison, $f_{CEO,S,k=0}$ and $f_{rep,S,k=0}$ are calibrated directly on the data under consideration by minimizing the deviation between our observed line positions and the tabulated values. The deviation between our measurement and HITRAN is plotted in figure 7 a) as grey dots. The red line indicates a linear fit to the data. A non-vanishing slope of the linear fit indicates an error in $f_{rep,S,k=0}$. Thus, $f_{rep,S,k=0}$ is chosen such as to minimize the slope of the linear fit. The offset of the linear fit is minimized by the value of $f_{CEO,S,k=0}$. Following this procedure, we find that all line positions agree with the HITRAN line positions to within 0.00012 cm$^{-1}$ (+/- 4 MHz, red shaded area in figure 7 a), with a standard deviation of 0.00006 cm$^{-1}$ (+/- 2 MHz). The one outlier at 1308.395 cm$^{-1}$ is most likely caused by a weak nearby transition that distorts the fit. From this, we conclude that the relative wavenumber axis generated in the QCL dual-comb measurement is accurate to within 0.00012 cm$^{-1}$ (+/- 4 MHz).

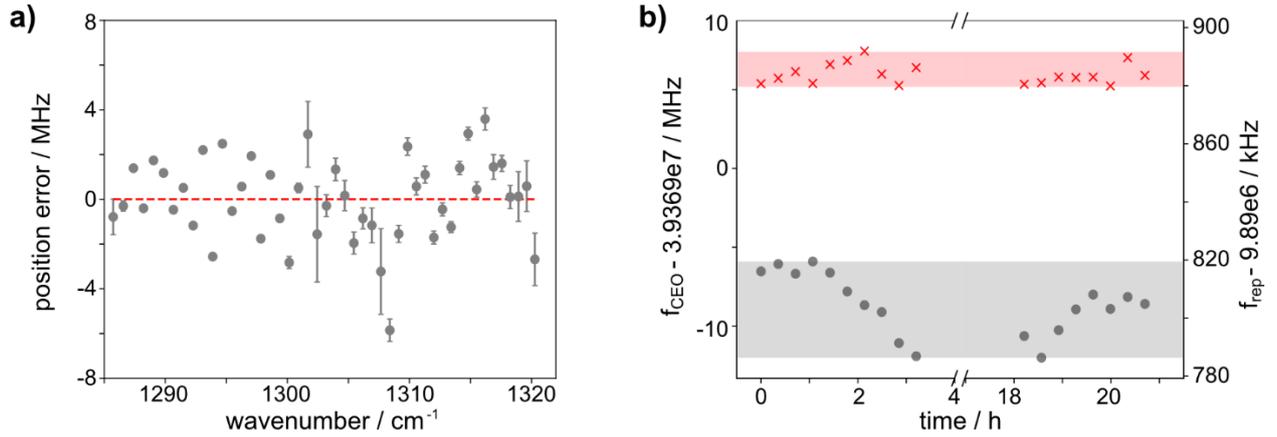

*Figure 7. a) Deviation between fitted line positions and tabulated values in Hitran (grey dots) for a measurement of $N_2O$ at 0.045 mbar. Error bars indicate the fit uncertainty. b) Evolution of the laser center frequency $f_{CEO,S,k=0}$ (grey dots, left axis) and repetition frequency $f_{rep,S,k=0}$ (red crosses, right axis) over 20 hours. The grey shaded area indicates a range of $f_{CEO}$ of 6 MHz and the red shaded area indicates a range of $f_{rep}$ of 12 kHz.*

Applying the same procedure to measurements of the $v_4$ band of methane, we have measured the position of 66 methane transitions (between 1283 and 1333 cm$^{-1}$) and compared them to HITRAN2020. IRis-F1 reproduces accurately the HITRAN data with a maximum deviation of < 0.0004 cm$^{-1}$ (12 MHz), which is within the stated uncertainties provided by HITRAN. A relative wavenumber accuracy very similar to that shown for $N_2O$ is found when comparing our data to data from Germann *et al.* [67] measured using a Fourier transform spectrometer based on a GPS stabilized frequency comb [65]. This comparison is to be confirmed once the used line list has been published and peer reviewed. It has to be pointed out that, while many of the methane transitions are located in the region having a larger level of noise, the agreement with tabulated positions remains very good.

The repeatability of the absolute frequency calibration of subsequent measurements is assessed by determining $f_{CEO,S,k=0}$ and $f_{rep,S,k=0}$ in a series of measurements spread over a time range of more than 20 hours. This should simulate an experiment starting with a calibration measurement and subsequent measurements of samples of interest are taken later the same day or on the following day. In figure 7b), we show the time evolution of $f_{CEO,S,k=0}$ and $f_{rep,S,k=0}$ over 20 hours. The peak-to-peak deviation of $f_{CEO,S,k=0}$ amounts to 6 MHz. For $f_{rep,S,k=0}$, we observe peak-to-peak fluctuations of 12 kHz. Multiplied by the maximum of 180 comb modes, this amounts to a maximum frequency error of ~2 MHz. Summing the two contributions of the absolute frequency drift with the 4 MHz maximum error of the relative frequency axis, we estimate an overall maximum frequency error of < 12 MHz (< 0.0004 cm$^{-1}$) of the dual-comb spectrometer.

### 3.3 Instrumental distortion

A key advantage of laser-based spectrometers is their high spectral resolution. The jitter of the laser emission frequency on a time scale of a single measurement will lead to a broadening of observed spectral features and thereby limit the resolving power of the spectrometer.

For this characterization, we recorded spectra of pure methane at low pressure (< 1 mbar). In these conditions, the contribution of the collisions to observed profile is negligible and the experimental profile is a convolution between a Gaussian which represents the Doppler broadening and the apparatus function. Indeed, the collisional broadening is very weak and is negligible under these experimental conditions. The well-known Doppler half-width can be calculated as

$$\gamma_D = \sqrt{\frac{2\ln(2)k_0 T}{mc^2}}\nu_0,$$

with $k_0$ the Boltzmann constant, T = 296 K the laboratory temperature, *m* the active molecule mass, *c* the speed of light and $\nu_0$ the center wavenumber of the transition.

We have tried to determine the width of the apparatus function and its line shape. We have tested a Gauss, Lorentz or Voigt profile, leading to Gauss or Voigt observed line shape (convolution between the Doppler line shape and the apparatus function). For each considered line, the low-pressure line shape was fitted with the three tested profiles. An example of the fit and residuals is given in Fig.8 for the P(3) $F_1(2)\leftarrow F_2(1)$ methane transition. The residuals do not show particular features, are very similar to each other and have the magnitude of the experimental noise. In Tab.1, the average values, over all studied lines, of the standard deviation of the residuals are presented. The three tested apparatus functions have very close and small values. The obtained half-widths of the apparatus functions are insignificant regarding the resolution of the spectrometer.

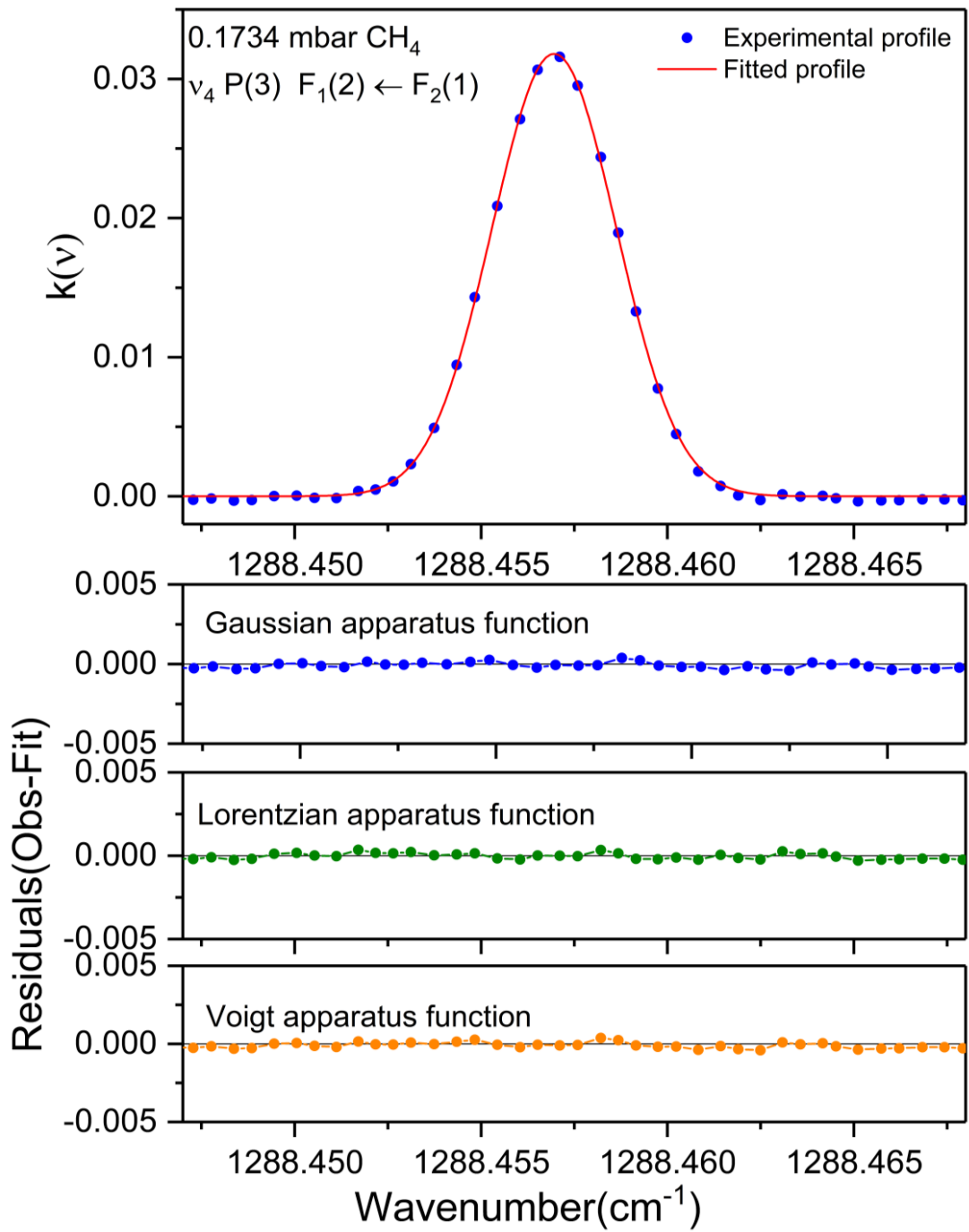

*Figure 8: Example of low-pressure observed (●) profile and the Gaussian fitted profiles for the P(3) $F_1(2) \leftarrow F_2(1)$ line in the $\nu_4$ band of $CH_4$. The residuals (Obs-fit) for considered Gauss, Lorentz and Voigt apparatus functions are shown at the bottom.*

|  | Gaussian apparatus function | Lorentzian apparatus function | Voigt apparatus function |
|---|---|---|---|
| Mean standard deviation of the residuals (cm$^{-1}$) | 0.00017 | 0.00017 | 0.00016 |

*Table 1: Average values, over all studied lines, of the standard deviations of the residuals (Exp-fit) for the Gaussian, Lorentzian and Voigt apparatus functions.*

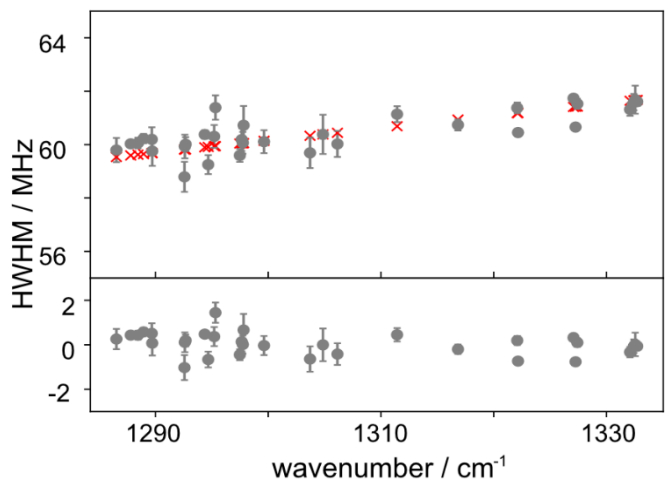

*Figure 9: Upper panel: Half width at half maximum (HWHM) of Gaussian line fits (grey dots) and calculated Doppler width (red crosses) for a measurement of CH$_4$ at 0.182 mbar. Lower panel: deviation between calculated and observed HWHM.*

Since none of the tested apparatus function gave significant improvement of the residuals, we compare the half-width at half-maximum of the fitted Gaussian profiles to the theoretical Doppler half-width at room temperature for methane lines within the IRis-F1 spectral range. Since the determination of the line width requires a higher SNR than determining the line center, we limit the analysis to the 31 transitions with SNR > 50. Figure 9 shows the difference between the fitted Gaussian (grey dots) and the theoretical Doppler half-widths (red crosses) as a function of their line positions. No systematic deviation of the observed absorption features was found. Rather, the observed line widths scatter around the theoretical Doppler values with a RMS deviation of 500 kHz (peak deviation < 1.5 MHz / 0.00005 cm$^{-1}$). We attribute the observed deviations to the remaining inaccuracy of the relative wavenumber axis and the limited SNR of the measurement. This result demonstrates that the line distortion due to the instrument response is negligible under the current experimental conditions (Doppler broadening and SNR) and that QCL frequency combs are well suited for applications that require high spectral resolution such as line-shape parameter measurements.

### 3.4 Example of results in high resolution molecular spectroscopy

We have recorded spectra of methane-nitrogen mixtures at 293K (±1 K) with an optical path length of 15.0 cm. The lines under study were located in the P-, Q and R-branches of the $\nu_4$ band of methane. The partial pressure of CH$_4$ was kept constant and very low (0.606 mbar), in view to minimize the self-broadening contribution, while the nitrogen pressure was comprised between 26.2 and 105.5 mbar (Fig. 10).

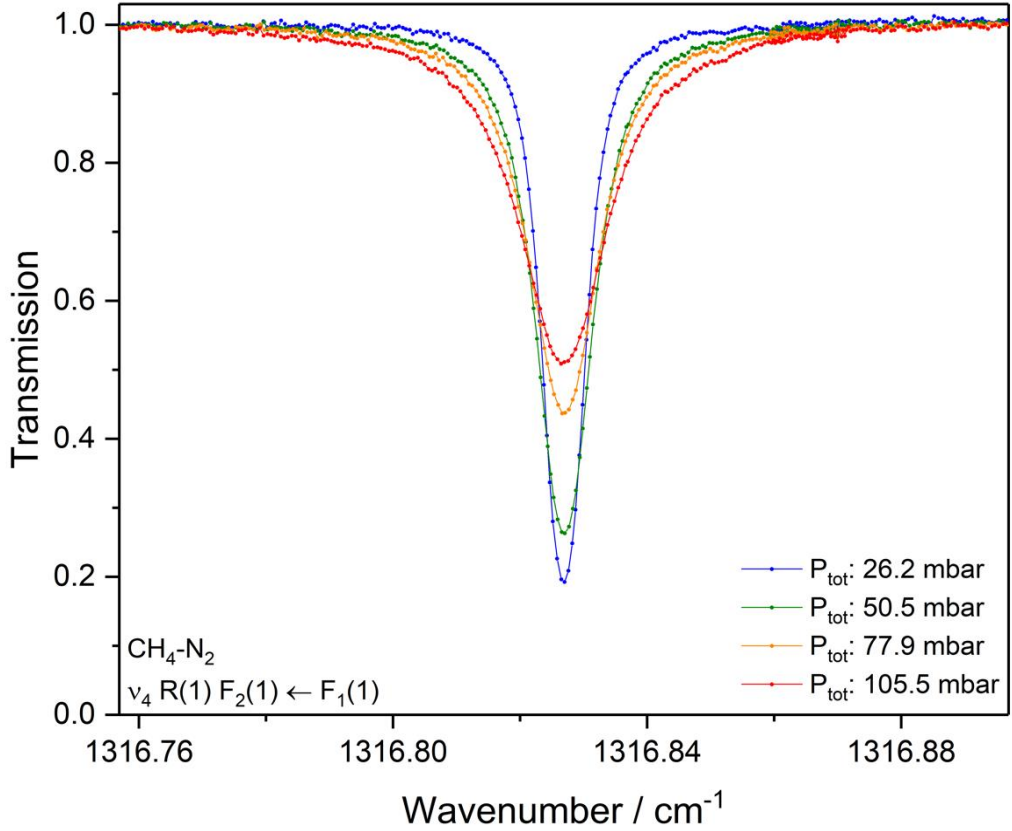

*Figure 10: Example of high-resolution IRis-F1 spectra of methane diluted in nitrogen for the ν4 band line R(1) F2(1) ← F1(1). The wavenumber axis calibration was performed using the described procedure. The blue, green, orange and red spectra were recorded with 0.606 mbar of methane diluted in 26.2, 50.5, 77.9 and 105.5 mbar of nitrogen, respectively.*

To determine the $N_2$-collisional width of lines, a procedure of fit was used. In order to compare with published data, we have considered the well-known Voigt profile [68] that is given by:

$$k_V(x, y, A) = A \frac{y}{\pi} \int_{-\infty}^{+\infty} \frac{exp(-t^2)}{y^2 + (x-t)^2} dt$$

with $A = \frac{S\sqrt{ln2}}{\gamma_D \sqrt{\pi}}$, $y = \sqrt{ln2}\frac{\gamma_C}{\gamma_D}$, $x = \sqrt{ln2}\frac{\nu-\nu_0-\delta_C}{\gamma_D}$, where $\nu_0$ is the line center position (in cm$^{-1}$), $S$ the line intensity (in cm$^{-2}$), $\gamma_C$ the collisional half-width (in cm$^{-1}$), $\delta_c$ the pressure-shift (in cm$^{-1}$) and $\gamma_D$ the calculated Doppler half-width (in cm$^{-1}$). The Voigt profile was adjusted on the experimental profile for each considered pressure of $N_2$ using the Levenberg-Marquardt algorithm [69]. Figure 11 gives an example of Voigt profile adjusted on the experimental line shape of the $\nu_4$ band R(1) F$_2$(1) ← F$_1$(1) transition. The pressure of methane was 0.606 mbar for a total pressure of 26.2 mbar. Measurements were performed at 293 K with an optical path length of 15.0 cm. Agreement between observed and adjusted profiles is shown by the residuals, which have the typical "W" shape signature for the Voigt profile. This expected residual signature shows that the spectrometer can also be used to study beyond-Voigt line shape parameters.

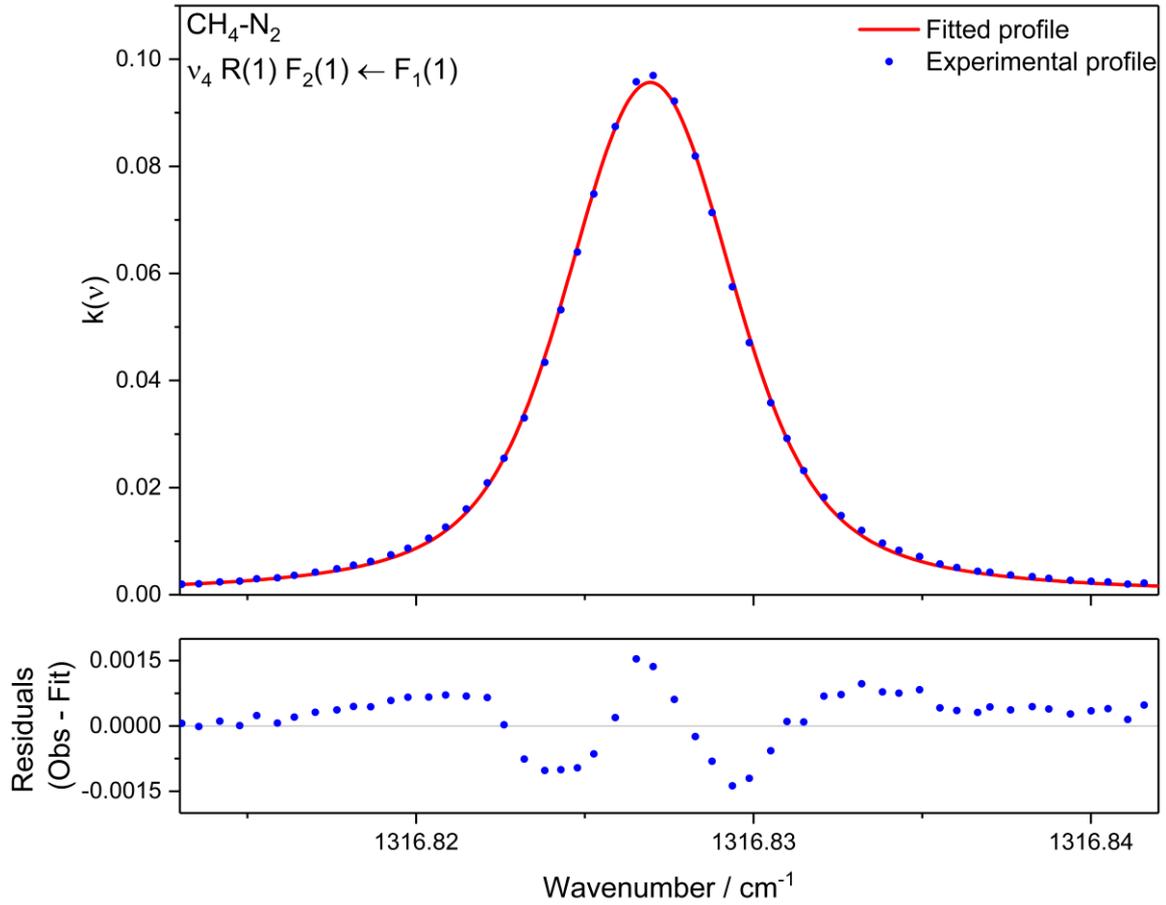

*Figure 11 Example of observed (●) and Voigt fitted profiles for the R(1) F2(1) ← F1(1) line in the ν4 band of CH4 for 0.606 mbar diluted in 26.2 mbar of N2 at 293K. The residuals (O-C) are shown at the bottom.*

The Voigt fits allow the measurement of $N_2$-collisional half-width of lines ($\gamma_C$ in cm$^{-1}$) at a given pressure of nitrogen. From these results, we can deduce the $N_2$-collisional broadening coefficient of lines ($\gamma_0$ in cm$^{-1}$.atm$^{-1}$) taking into account the weak self-broadening:

$$\gamma_C = \gamma_0 \times p_{N_2} + \gamma_0^{Self} \times p_{CH_4}$$

where $p_{N_2}$ and $p_{CH_4}$ are the $N_2$- and $CH_4$-pressures, while the self-broadening coefficients of lines ($\gamma_0^{Self}$ in cm$^{-1}$.atm$^{-1}$) are taken from Ref [63]. For that, $N_2$-broadening coefficient is given by the slope of the best-fit straight line passing through the fitted half-widths at different pressures. An example is shown in Fig. 12.

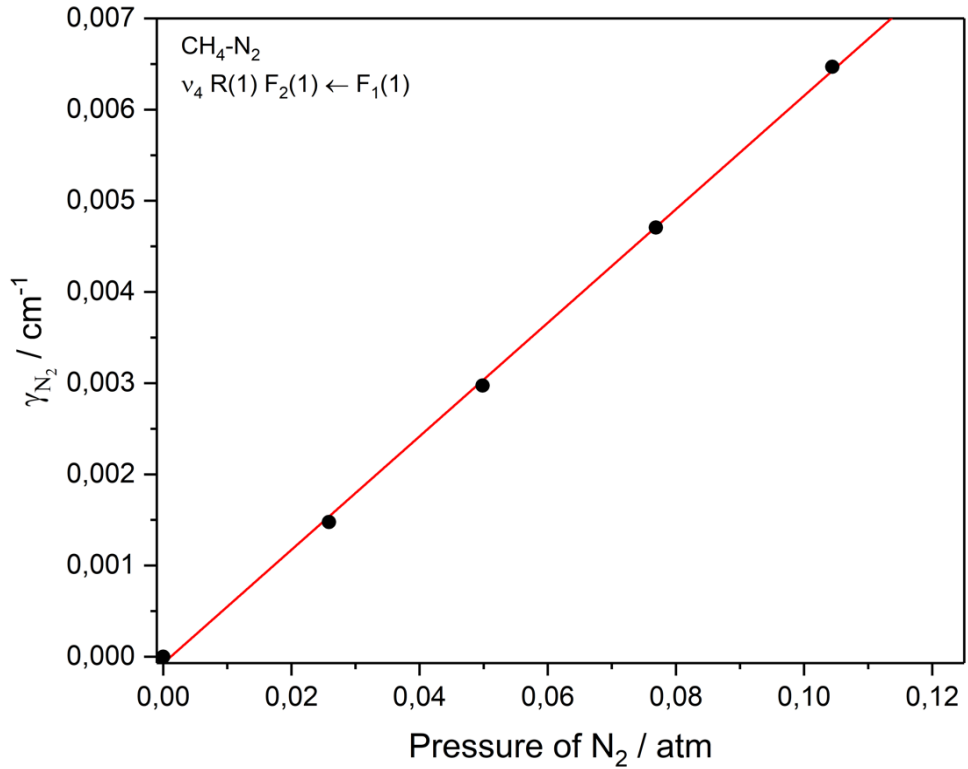

*Figure 12 Plot of the nitrogen collisional half-widths γc-γ0self*pCH4 versus the N2-pressure for the R(1) F2(1) ← F1(1) line of the ν4 band of methane. The half-widths are derived from the fits of the Voigt (□) profile. The slope of the best-fit line, passing through zero, represents the N2-broadening coefficient γ0. The γ0self of CH4 data are taken from [63].*

40 transitions were analyzed using the described procedure. The uncertainties are determined as twice the standard deviation of the linear regression plus 2% of the coefficient itself in order to take into account the experimental errors such as the pressure, the optical path length measurements, or the considered line shape. The results are presented in Fig. 13 where we compare with previous studies in the $\nu_4$ band of methane. The work of Devi *et al.* [70] were performed using a tunable diode-laser and a Fourier transform spectrometers with Voigt fits; Varanasi *et al.* [71] used a tunable diode-laser spectrometer and consider a Voigt profile; Smith *et al.* [72] used a Voigt profile on FTIR line-shape measurements. Finally, Lepère *et al.* [73] and Martin *et al.* [74] used a tunable diode-laser spectrometer to perform $N_2$-broadening measurements considering the Voigt profile as well as line-shape models, which take into account fine physical effects. Figure 13 separates the A, E and F-species lines. In general, the lines under study are in good agreement with the literature, considering our experimental uncertainties (which are very conservative).

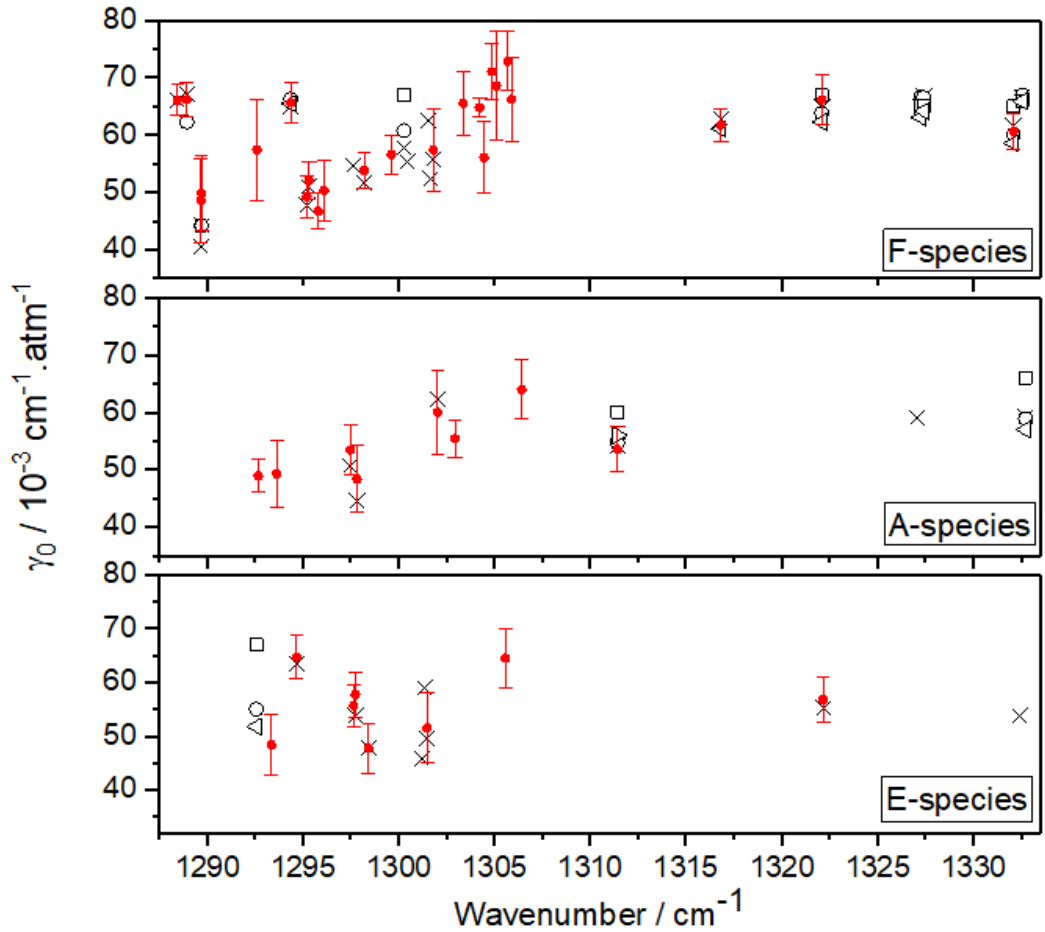

*Figure 13 Comparison between our results (●) and previous measurements for the same transitions in the $\nu_4$ band: (○) Devi et al. [70] (TDLS and FTIR), (□) Varanasi et al. [71] (TDLS), (×) Smith et al. [72] (FTIR), (▷) Lepère et al. [73] (TDLS), and (◁) Martin et al. [74] (TDLS). The results are separated for each symmetry species (A, E and F).*

## 4. CONCLUSION AND PERSPECTIVES

The IRis-F1 is a tabletop dual-comb spectrometer building on quantum cascade laser frequency combs and emitting in the mid-infrared. We hereby report on the development of the *step-sweep* mode, which allows for high-resolution measurements with a very accurate knowledge of the wavenumber axis. Despite the optical and electronic simplicity of the presented technique, we demonstrate that we are able to record complete spectra covering over 50 $cm^{-1}$ in 30 minutes with a frequency accuracy of below 12 MHz (0.0004 $cm^{-1}$) in separately calibrated measurements and below 4 MHz (0.00012 $cm^{-1}$) when the calibration is simultaneous with the sample measurement. Furthermore, measurements of Doppler broadened spectra have shown that the line distortion due to the instrument response is negligible. With these characteristics and the transmission noise floor of down to $10^{-3}$, the *step-sweep* technique can be employed for challenging studies in molecular spectroscopy. This is illustrated by $N_2$-collisional broadening measurements in the $\nu_4$ band of methane lines. We envisage to use the spectrometer for a range of other studies such as the precise measurement of individual line intensities or line-mixing effect, including other spectral ranges between 900 $cm^{-1}$ to 2300 $cm^{-1}$ accessible with QCL frequency combs.

## ACKNOWLEDGEMENTS

We acknowledge A. Foltynowicz for sharing her unpublished $CH_4$ line list to support our analysis. M. Lepère is acknowledging support from "Fonds de la Recherche Scientifique (F.R.S.-FNRS)". B. Vispoel acknowledges the F.R.S-FNRS for the post-doctoral support. IRsweep acknowledges funding for this project from the European Union's Horizon 2020 research and innovation programme under the Marie Skłodowska-Curie grant agreement No 101032761.